\documentclass{article}

\usepackage{amsmath}
\usepackage{graphicx}
\usepackage[numbers]{natbib}
\usepackage[caption=false]{subfig}

\begin{document}
\title{Matched transient and steady-state approximations of first-passage-time distributions of coloured noise driven leaky neurons}
\author{Akke Mats Houben}
\date{\today}
\maketitle

\begin{abstract}
The first-passage-time distribution of a leaky integrate-and-fire neuron driven by a characteristically coloured noise is approximated by matching a transient and a steady-state solution of the membrane voltage distribution.
These approximations follow from a simple manipulation, made possible by the specific `eigen' colouring of the noise, which allows to express the membrane potential as a Gaussian diffusion process on top of a deterministic exponential movement.
Following, the presented method is extended to the case of an arbitrarily coloured noise driving by factoring out the `eigen' noise and replacing the residue with an equivalent Gaussian process.
It is shown that the obtained expressions agree well with numerical simulations for different values of the neuron parameters and noise colouring.
\end{abstract}

\section{Introduction}
Because neurons in the brain receive noisy inputs and generate irregular output spike trains, a lot of research effort is devoted to understanding the spiking statistics of neuronal models incorporating some noise. Since an important measurement of neuronal activity is the rate of firing of a neuron, particular interest is taken in the investigation of the inter-spike-interval (being the inverse of the firing-rate) statistics of the irregular spiking output of neurons receiving noisy input.

\begin{figure}\centering
    \includegraphics[width=0.5\textwidth]{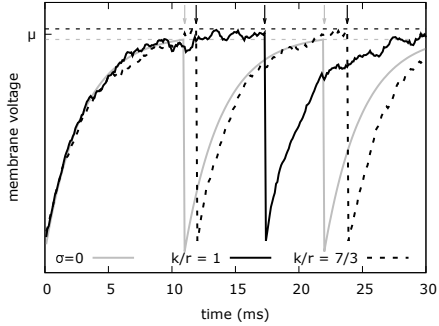}
    \caption{\textbf{Simulated membrane potential} of leaky integrate-and-fire neurons driven by a constant input (grey line) and noise inputs (black lines) with different colouring (indicated by line-type). Firing times are indicated with arrows.}\label{fig:vs}
\end{figure}

An important group of models for the study of spiking activity is the set of integrate-and-fire (IF) neurons \citep{Brunel2007, Burkitt2006a, Lapicque1907}, which consist minimally of a description of the evolution of a membrane potential $v$, which, as suggested by the name, describes the integration of some input, combined
with a spiking mechanism of some sort. A typical choice is resetting the membrane potential to some reset value $v_0$ once it reaches a threshold value $v_c$, after which $v$ follows again the description of the sub-threshold evolution, but other mechanisms to account for spiking can be used.
A widely used variant is the leaky integrate-and-fire (LIF) neuron:
\begin{equation}\label{eq:dv}
    \frac{1}{k}\frac{dv}{dt} = I(t) - v(t),
\end{equation}
which integrates an input $I(t)$, and for which, importantly, after a perturbation the voltage across the membrane decays exponentially back to zero with a time-constant $1/k$. When a constant input $I(t) = \mu > v_0$ is applied, the voltage across the membrane follows an exponential approach to the value $v_{\infty} = \mu$, from a starting potential $v_0$. Using the spike-and-reset mechanism as described before ($v > v_c \implies v\leftarrow v_0$) the state of the neuron is completely reset after spiking, so the evolution of the membrane potential during each inter-spike-interval is independent from the other intervals.
For a given constant input above the firing threshold $\mu>v_c$, the lenght of the inter-spike-interval $\tau_{isi}$ is equivalent to the time it takes for the membrane potential to reach the threshold value $v_c$ starting from the reset value $v_0$:
\begin{equation}
    \tau_{isi} = -\frac{1}{k}ln\left(\frac{v_c-\mu}{v_0-\mu}\right). \nonumber
\end{equation}
Fig. \ref{fig:vs} shows the deterministic evolution of the membrane voltage of the LIF neuron (grey line) driven by a supra-threshold constant input. The asymptotic value of $v$ lies above the firing-threshold, so the neuron spikes tonically with a constant inter-spike-interval, as indicated in the figure.

However, neurons typically show irregular spiking behavior, which arises due to a range of noise sources \citep[e.g.][]{Destexhe2012, Ermentrout2008, Faisal2008, Holden1976, Stevens1972, Tuckwell1989} such as channel noise,
%\citep[e.g.][]{Chow1996, White2000, }
stochastic synaptic transmission, %\citep[e.g.][]{cite},
transmission and spiking failures, %\citep[e.g.][]{cite},
and fluctuations due to large numbers of simultaneous excitatory and inhibitory inputs. % \citep[e.g.][]{vanVreeswijk1996, vanVreeswijk1998, }. 
Grouping the different noise sources together, an easy approximation is to consider an input consisting of a mean drive with an added noise $\eta(t)$:
\begin{equation}\label{eq:input}
  I(t) = \mu + \sigma\eta(t),
\end{equation}
in which case the firing of the neuron is characterised by an inter-spike-interval distribution. A closely related problem is to find the first-passage-time distribution which, in case the state of the neuron is completely reset after a spike and the neuron starts from the reset potential $v(0)=v_0$, is equal to the inter-spike-interval distribution.

The (first two) moments of the inter-spike-interval distribution have reported exact expressions for noise driven leaky \citep{Capocelli1971, Johannesma1968, Ricciardi1979, Roy1969}
 and perfect\footnote{an IF neuron that integrates input without a drive back to a resting potential, i.e. Eq. \ref{eq:dv} without the subtraction of $v$ on the r.h.s} \citep{Holden1976, Tuckwell1988} IF neurons, and in case of several extensions to the leaky IF neuron model such as the incorporation of an exponentially decaying threshold \citep{Lindner2005} or models in which the spike variability is given by a stochastic threshold \citep{Braun2015}. An exact expression for the full inter-spike-interval distribution of the perfect integrate-and-fire neuron under white noise driving has long been known \citep{Gerstein1964}, and has been extended to long range correlated \citep{Middleton2003} and coloured \citep{Lindner2004} noise driving.
 Expressions for the full inter-spike-interval or first-passage-time distribution of the noise driven leaky IF neuron have been found analytically for restricted sets of neuron parameters \citep{Siebert1969, Sugiyama1970}, for long-time correlated noise by a quasi-static approximation \citep{Swalger2008}, or in Laplace transformed form \citep{Capocelli1971, Roy1969, Sugiyama1970}. 

 This letter presents another approximation of the first-passage-time distribution of a leaky IF neuron with weak restrictions on the neuron parameters. The derivation of the expression becomes apparent by driving the neuron with a specific noise, filtered with a time-constant that matches the membrane time-constant. 
 Using this `eigen' noise driving, the drift term is replaced by a deterministic exponential transient and the steady-state reduces to a Brownian motion, thus making it possible to obtain approximations of the membrane potential distribution on a short and long timescale, which can be matched on an intermediate time-scale. 
 It is then shown that using this specific approximation, it is possible to obtain expressions for arbitrary noise colouring, by `factoring' out the characteristic noise and replacing the residual noise with an equivalent Gaussian process.
 
\section{Specifically `eigen' coloured noise}\label{sec:eigennoise}
By using as the noise term $\eta(t)$ of (\ref{eq:input}) the sum of a Gaussian white noise $dW/dt$ and a Wiener process $rW$, being the integral of $dW/dt$ multiplied by a constant $r$, thus
\begin{equation}
    I(t) = \mu + \sigma\left(dW/dt + rW\right), 
\end{equation}
a manipulation to (\ref{eq:dv}) will be possible that makes apparent an approximation of the first-passage-time distribution.
Fig. \ref{fig:vs} shows the evolution of the membrane potential of some neurons driven by this coloured noise (black lines), for different values for $r$ (indicated by different line types). 
Subtracting the mean drive $\mu$ from $I(t)$, and defining $\sigma$ to be the noise standard deviation multiplied by $k$; $x := v-\mu$; $x_0:=v_0-\mu$ and; $x_c:=v_c-\mu$ for notational convenience, allows the sub-threshold membrane equation to be written as a stochastic differential equation
\begin{equation}\label{eq:dx}
    dx = -kxdt + \sigma\left[ r Wdt + dW \right], 
\end{equation}
with its Laplace transform:
\begin{equation}\label{eq:dxL}
    (k+s)\widetilde{x} - x_0 = \sigma(r+s)\widetilde{W},
\end{equation}
where it is assumed that $W(0)=0$.
This section will consider the specific case in which the constant $r$ multiplying the Wiener process is equal to the inverse of the time-constant of the membrane, thus $r = k$. 
Section \ref{sec:arbitraryNoise} will extend the method developed here to the case of arbitrary noise colouring, obtained by any $r\geq0$. % perhaps r\in R? is r restricted?

In the case that $r=k$ dividing both sides of (\ref{eq:dxL}) by $(k+s)$ gives
\begin{equation}
    \widetilde{x}  = \sigma \widetilde{W} + \frac{x_0}{k+s}. \nonumber
\end{equation}
Then by applying the inverse Laplace transform we see that the dynamics of $x$ are those of a Brownian motion (or white noise driven perfect IF neuron \citep{ Gerstein1964}), for which exact expressions for the first-passage-time distributions are known \citep{Gerstein1964, Schrodinger1915, Tuckwell1988}, on top of an exponential drive:
\begin{equation}\label{eq:dy}
    x = \sigma W + x_0e^{-kt}.
\end{equation}

\subsection{Probability density and first-passage-time distribution}
The density of $x$ in (\ref{eq:dy}) is given by the Fokker-Planck equation
\begin{equation}\label{eq:fpex}
    \frac{\partial}{\partial t}p(x,t) = kx_0e^{-kt}\frac{\partial}{\partial x}p(x,t) + \frac{\sigma^2}{2}\frac{\partial^2}{\partial x^2}p(x,t),
\end{equation}
with an absorbing boundary $p(x_c,t) = 0$ at the spiking threshold $x_c$ and initial condition $p(x,0) = \delta(x-x_0)$.

For sufficiently fast neurons \textemdash large enough $k$\textemdash the drift term only influences the solution for small $t$.
If, additionally, the firing threshold is sufficiently far from the fixed point $x=0$, the solution does not interact with the boundary for small $t$. So the transient solution $p_k(x,t)$ to (\ref{eq:fpex}) can be approximated by ignoring the absorbing boundary:
\begin{equation}\label{eq:p_k}
    p_k(x,t) = \frac{1}{\sqrt{2\pi \sigma^2 t}} \exp{\left( -\frac{(x-x_0 e^{-kt})^2}{2\sigma^2 t} \right)}.
\end{equation}
Conversely, the drift term can be assumed to be vanishing for large $t$, so the steady-state solution $p_1$ is unaffected by the drift, leading to the approximation 
\begin{equation}\label{eq:p_1}
    p_1(x,t) = \frac{1}{\sqrt{2\pi\sigma^2t}}\left[\exp{\left(-\frac{x^2}{2\sigma^2t} \right)} - \exp{\left(-\frac{(x-2x_c)^2}{2\sigma^2t} \right)} \right],
\end{equation}
using the method of images.

To match the two solutions we require them to overlap in an intermediate time-scale. 
For the transient solution (\ref{eq:p_k}) it is easy to see that the solution tends to a Brownian motion with zero mean and variance $\sigma^2/2 t$ for large $t$. For the steady-state (\ref{eq:p_1}), with sufficient distance between the threshold $x=x_c$ and fixed point $x=0$, we only have to consider the solution in the domain $x\in(-\infty, x_c]$. This leads to the two solutions to be matched in the limits:
\begin{align}
    p_o(x,t) &= \lim_{t\to \infty} p_k(x,t) = \lim_{t\to 0} p_1(x,t) \nonumber \\
    &= \frac{1}{\sqrt{2\pi\sigma^2t}}\exp{\left( - \frac{x^2}{2\sigma^2t} \right)}, \nonumber
\end{align}
and results in the complete matched approximation
\begin{align}\label{eq:p_c}
    p(x,t) &= p_k(x,t) + p_1(x,t) - p_o(x,t) \nonumber\\
    &= \frac{1}{\sqrt{2\pi\sigma^2t}}\left[\exp{\left(-\frac{(x-x_0e^{-kt})^2}{2\sigma^2t} \right)} - \exp{\left(-\frac{(x-2x_c)^2}{2\sigma^2t} \right)} \right].
\end{align}

The matched solution will incur an error proportional to
\begin{equation}
    \epsilon = -x_0 \exp{\left(-k \tau_1\right)}, \text{ with } \tau_1 = \frac{x_c^2}{3\sigma^2}, \nonumber \\
\end{equation}
where $\tau_1$ is the typical time that the drift-less image starting at $x=2x_c$ crosses the threshold $x=x_c$ from above, and thus enters the domain $x\in(-\infty,x_c]$. The error is then proportional to $\epsilon$, the distance of the mean of the transient solution to the fixed-point $x=0$ at the typical time the drift-less image enters the solution domain. 
From the definition of $\epsilon$ and $\tau_1$, one can see that the error should be low for fast neurons ($k$ large) and for long typical passage times $\tau_1$, which occurs for large ratios between the firing threshold $x=x_c$ and the diffusion coefficient $\sigma^2/2$. The error is also linearly influenced by the distance of the reset potential $x=x_0$ to the fixed-point $x=0$.

Integrating (\ref{eq:p_c}) over the possible values for $x$ gives the approximated probability distribution that at time $t$ a neuron has not fired yet,
\begin{align}\label{eq:S}
    S(t) &= \int_{-\infty}^{x_c} p_c(x,t)dx \nonumber \\
    &= \frac{1}{2}\left[erf{\left( \frac{x_c-x_0e^{-kt}}{\sqrt{2\sigma^2t}}\right)} + erf{\left(\frac{x_c}{\sqrt{2\sigma^2t}} \right)} \right],
\end{align}
and $1-S(t)$, the probability that a neuron has fired at some time before $t$. Changes in $1-S(t)$ with respect to time then relate to the probability of the timings of threshold crossings. The approximated first-passage-time distribution is thus the time-derivative of $1-S(t)$:
\begin{align}\label{eq:fpt}
    f(t) =&~ \frac{\partial}{\partial t} [1-S(t)] \nonumber \\
    =&~ \left[\frac{x_c-x_0e^{-kt}}{2\sqrt{2\pi\sigma^2 t^3}} - \frac{x_0 k e^{-kt}}{\sqrt{2\pi\sigma^2 t}} \right]\exp{\left(-\frac{(x_c-x_0 e^{-kt})^2}{2\sigma^2t} \right)} \nonumber \\
    &+ \frac{x_c}{2\sqrt{2\pi\sigma^2t^3}}\exp{\left(-\frac{x_c^2}{2\sigma^2t} \right)}.
\end{align}

\begin{figure}
    \centering
    \includegraphics[width=0.5\textwidth]{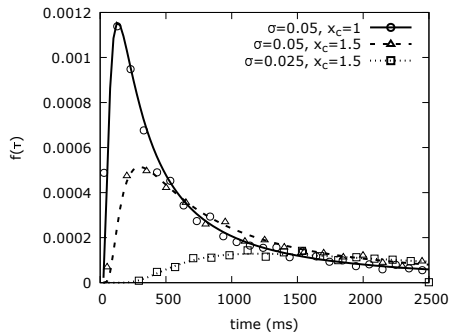}
    \caption{\textbf{First-passage-time distributions} obtained by numerical simulation of (\ref{eq:dx}) (scatter plots) together with the theoretical approximation $f(t)$ (lines) for some different parameters (indicated by different line types). The parameter values, besides those mentioned in the legend, used are $k=1/5$ and $x_r=-3$. The numerical results are from $20000$ neurons per line simulated for $25000$ time-steps of size $\Delta t=0.1$. }\label{fig:fpt}
\end{figure}

Fig. \ref{fig:fpt} shows numerically measured first-passage-time distributions alongside $f(t)$ for some different parameters. 

It is clear that the obtained expressions (\ref{eq:p_c}), (\ref{eq:S}) and (\ref{eq:fpt}) reduce to the exact expressions of the perfect integrate-and-fire neuron \citep{Gerstein1964} when $x_0=0$. However, this is not the case when $k\to0$. In this case the time-constant approaches infinity, leading to no changes in the membrane potential. This can also be inferred from (\ref{eq:dv}), which can be re-written as
\begin{equation}
    \frac{dv}{dt} = kI(t) - kx(t), \nonumber
\end{equation}
with both the input and the `leak' term multiplied by $k$, resulting in no net change of the membrane potential for $k=0$.

\section{Extension to arbitrarily coloured noise}\label{sec:arbitraryNoise}
The method described in section \ref{sec:eigennoise} results in an approximate expression for the first-passage-time density in a very restricted case: it demands that the input noise is filtered with a time-constant that matches the time-constant of the membrane equation. To be of real interest, however, a solution needs to exist for a wider range of input parameters. In the following, it is shown how this method can be extended to arbitrary noise driving, with $r>0$, by utilising the Gaussian nature of the solution.

Eq. (\ref{eq:p_c}) describes the superposition of two Gaussian processes with equal covariances $\sigma^2 t$, but differing means $\mu_k = x_0e^{-kt}$ and $\mu_1 = 2x_c$, and opposite signs $p_k>0$ and $(p_1-p_o)<0$. Since Gaussian processes are characterised completely by their mean and covariance, this information is enough to reconstruct the solution $p_c(x,t)$ for $r=k$. Solutions for $r\neq k$ are expected to similarly be a superposition of Gaussian processes, the aim is then to obtain the mean and variance of the membrane potential distribution and replace the means and variances of $p_c$ with the obtained expressions.

Starting from the SDE
\begin{equation}
    dx = -kxdt + \sigma\left[ r Wdt + dW \right], \nonumber
\end{equation}
and following the same method as before:
\begin{align}
    (k+s)\widetilde{x} - x_0 &= \sigma(r+s)\widetilde{W}  \nonumber \\
    \widetilde{x} &= \sigma \frac{r+s}{k+s}\widetilde{W} +  \frac{x_0}{k+s} \nonumber \\
        &= \sigma\left[1+\frac{r-k}{k+s}\right]\widetilde{W} +  \frac{x_0}{k+s}, \nonumber
\end{align}
leads to $x$ being given by:
\begin{equation}\label{eq:dxc}
    x = \sigma\left[W_t + (r-k)\int_0^{t}e^{-k(t-\tau)}W_\tau d\tau\right] + x_0e^{-kt},
\end{equation}
a Gaussian process with mean 
\begin{align}
    E\left[x\right] =&~ \sigma E\left[W(t)\right] + \sigma E\left[ (r-k)\int_0^t e^{-k(t-\tau)}W(\tau) d\tau \right] + x_0e^{-kt} \nonumber \\
    =&~x_0e^{-kt}, \nonumber
\end{align}
and variance:
\begin{align}\label{eq:var}
    E\left[(x-E[x])^2\right] =&~ \sigma^2 E\left[(W(t)-E[W(t)])^2\right] \nonumber \\
    &+ \sigma^2(r-k)^2 E\left[\left( \int_0^t e^{-k(t-\tau)}W(\tau) d\tau - E\left[ \int_0^t e^{-k(t-\tau)}W(\tau)d\tau \right] \right)^2\right] \nonumber \\
    &+ \sigma^2 2 Cov\left[W(t),(r-k)\int_0^t e^{-k(t-\tau)}W(\tau)d\tau\right] \nonumber \\
    =&~ \sigma^2 \left[ t 
    + \frac{(r-k)^2}{k^2}\left( t + 2\frac{e^{-kt}-1}{k} - \frac{e^{-2kt}-1}{2k}\right ) 
    +  2\frac{r-k}{k}\left(t + \frac{e^{-kt}-1}{k} \right) \right].
\end{align}

Replacing the means and variances in (\ref{eq:p_c}) leads to the approximated probability distribution of $x$ for $r\neq k$
\begin{equation}
    p_{r\neq k}(x,t) = \frac{1}{\sqrt{2\pi \nu(t)}}\left[\exp{\left(-\frac{(x-x_0e^{-kt})^2}{2\nu(t)}\right)} - \exp{\left(-\frac{(x-2x_0)^2}{2\nu(t)} \right)}\right],
\end{equation}
with $\nu(t)$ being the expression of the variance obtained in (\ref{eq:var}).

\begin{figure}
    \centering
    \includegraphics[width=0.5\textwidth]{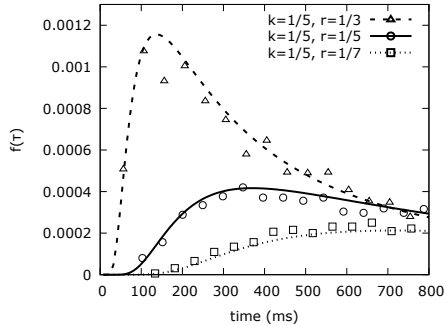}
    \caption{\textbf{First-passage-time distributions} for different noise colouring obtained by numerical simulation of (\ref{eq:dxc}) (scatter plots) together with the theoretical approximation $f_{r\neq k}(t)$ (lines) for some different parameters (indicated by different line types). Parameter values not mentioned in the legend are $x_c=1, x_r=-3$ and $\sigma=0.05$. The numerical results are from $20000$ neurons per line simulated for $25000$ time-steps of size $\Delta t=0.1$}
    \label{fig:fpt_c}
\end{figure}

The resulting expression $p_{r\neq k}(x,t)$ can similarly be integrated with respect to $x$
\begin{align}
    S_{r\neq k}(t) =&~ \int_{-\infty}^{x_c} p_{r\neq k}(x,t) dx \nonumber \\
    =&~ \frac{1}{2}\left[ erf\left(\frac{x_c -x_0e^{-kt}}{\sqrt{2\nu(t)}}\right) + erf\left(\frac{x_c}{\sqrt{2\nu(t)}} \right)\right]],
\end{align}
since $\nu(t)$ is independent of $x$.
Then again the approximation of the probability of a neuron described by (\ref{eq:dxc}) to have crossed the threshold $x_c$ before time $t$, $1-S_{r\neq k}(t)$, can be differentiated to obtain the approximate first-passage-time distribution 
\begin{align}\label{eq:fptc}
    f_{r\neq k}(t) =&~ \left[\frac{(x_c - x_0e^{-kt})\nu'(t)}{2\sqrt{2\pi\nu^3(t)}} + \frac{kx_0e^{-kt}}{\sqrt{2\pi\nu(t)}}  \right]\exp{\left(-\frac{(x_c-x_0e^{-kt})^2}{2\nu(t)}\right)} \nonumber \\
    &+ \frac{x_c \nu'(t)}{2\sqrt{2\pi \nu^3(t)}} \exp{\left( -\frac{x_c^2}{2\nu(t)}\right)},
\end{align}
with $\nu(t)$ being the variance (\ref{eq:var}) and 
\begin{equation}
    \nu'(t) = \sigma^2 \left[ 1 + \frac{(r-k)^2}{k^2}\left(1 - 2e^{-kt} + e^{-2kt} \right) + 2\frac{r-k}{k}\left(1 - e^{-kt}\right)\right], \nonumber
\end{equation}
the derivative of $\nu(t)$.
Notice that for $r=k$ the variance and its derivative reduce to $\nu(t)=\sigma^2t$ and $\nu'=\sigma^2$, respectively, so expression (\ref{eq:fpt}) is recovered from (\ref{eq:fptc}).
Fig. \ref{fig:fpt_c} shows the comparison between numerically obtained first-passage-times (scatter plots) and the theoretical $f_{r\neq k}(t)$ (lines), for different ratios $r/k$ (indicated by different line-types).

\section{Discussion}
This letter describes an approximate solution to the first-passage-time distribution of a leaky integrate-and-fire neuron for a characteristic noise-drive in which the filtering time-constant of the noise matches the time-constant of the neuron. Using this `eigen' noise driving, the diffusion equation reduces to a Brownian motion on a deterministic exponential, thus making it possible to obtain approximate solutions on two different time-scales of the membrane potential distribution that can be matched on an intermediate scale.
From this approximation then the approximate first-passage-time distribution can be constructed.
Furthermore, it is shown that the developed approximation also holds for noise colouring with different filtering constants.

A remaining question is how the low-pass filtered white noise emerges in the input, if it is assumed that the aggregated activity of a large group of neurons resembles a white noise. It has been shown that the extra-cellular environment of neurons has a low-pass filtering effect \citep{Bedard2004, Bedard2006}, and that the local-field-potential most strongly results from the activity of nearby neurons \citep{Destexhe1999, Katzner2009, Pettersen2008}. Thus a local population of independently firing neurons could account for the white noise $dW/dt$ through direct afferent inputs, whereas the correlated low-pass filtered component $W(t)$ is received through a local-field coupling.
%Interestingly, with regards to the characteristic noise, it has been shown that the addition of spike-trains of independent neurons with similar power-spectral-densities, leads the accumulated output to show the same power spectral density as the single neurons \citep{Lindner2006}. Thus, since the ISI of a neuron is influenced by its time constant $1/k$, a neuron within a population of sufficiently similar neurons will receive an input that matches its own time-constant, thus giving rise to the characteristic `eigen' noise driving used in section \ref{sec:eigennoise}.

Alternatively, the filtering could be due to internal processes or properties of parts of the neuron, such as (passive) filtering on dendrites \citep{Linden2010}. 
This would be interesting in light of the independence of subsequent inter-spike-intervals. For each inter-spike-interval to be independent would require that for a neuron spiking at time $t = t_{spike}$, in addition to a full reset of the state of the neuron $v(t_{spike})=v_0$, the accumulated noise is reset $W(t_{spike})=0$. In general a neuron has no such instantaneous influence on its external inputs, thus this requirement seems unlikely for an exogenously low-pass filtered noise component. However if the low-pass filtering effect is internal to the neuron, and in addition this accumulative effect is reset by a spiking event, expression (\ref{eq:fptc}) will give an approximation of the inter-spike-interval distribution of the neuron.

\bibliographystyle{plainnat}
\bibliography{article.bbl}%{library.bib}
\end{document}